\documentstyle[prl,aps,multicol]{revtex}
\renewcommand{\narrowtext}{\begin{multicols}{2} \global\columnwidth20.5pc}
\renewcommand{\widetext}{\end{multicols} \global\columnwidth42.5pc}
\begin{document}
\draft
\title{Eigenfunctions of electrons in weakly disordered quantum dots: \\
Crossover between orthogonal and unitary symmetries}
\author{{\bf E. Kanzieper and V. Freilikher}}
\address{The Jack and Pearl Resnick Institute of Advanced Technology,\\
Department of Physics, Bar-Ilan University, Ramat-Gan 52900, Israel}
\date{September 9, 1996}
\maketitle

\begin{abstract}
A one-parameter random matrix model is proposed for describing the
statistics of the local amplitudes and phases of electron eigenfunctions in
a mesoscopic quantum dot in an arbitrary magnetic field. Comparison of the
statistics obtained with recent results derived from first principles within
the framework of supersymmetry technique allows to identify a transition
parameter with real microscopic characteristics of the problem. The
random-matrix model is applied to the statistics of the height of the
resonance conductance of a quantum dot in the regime of the crossover
between orthogonal and unitary symmetry classes.
\end{abstract}

\pacs{\tt cond-mat/9609081}

\narrowtext

\section{Introduction}

Statistical properties of the electron eigenfunctions in disordered
quantum
dots have recently become a subject attracting considerable theoretical and
experimental interest \cite{FM} - \cite{PR1995}. One of the reasons is that
the problem of particle motion in a bounded disordered potential comprises a
particular case of general chaotic systems, such as quantum billiards \cite
{Argaman}. On the other hand, the development of the ``microwave-cavity''
technique \cite{Prigodin-et} - \cite{Stein} has opened up a unique
possibility for experimental studies of wave-function statistics.
Furthermore, the mesoscopic conductance fluctuations in a quantum dot in the
Coulomb blockade regime contain fingerprints of the statistics of the
eigenfunctions of confined electrons \cite{FE}, \cite{JSA} - \cite{Alhassid}%
. This also provides experimental access \cite{Chang,Folk} to the
wave-function statistics.

In order to characterize the eigenfunction statistics, we introduce two
distribution functions. The first refers to the local {\em amplitudes} of
electron eigenfunctions $\psi _n$ inside a dot of volume $V$, 
\begin{equation}
{\cal P}\left( v\right) =\left\langle \delta \left( v-V\left| \psi _n\left( 
{\bf r}\right) \right| ^2\right) \right\rangle \text{,}  \label{eq.01}
\end{equation}
while the second distribution function 
\begin{equation}
{\cal Q}\left( s\right) =\left\langle \delta \left( s-\sin \varphi _n\left( 
{\bf r}\right) \right) \right\rangle  \label{eq.02}
\end{equation}
is associated with the eigenfunction {\em phases}, $\varphi _n=\arg \psi _n$%
. Angular brackets $\left\langle ...\right\rangle $ denote averaging over
bulk impurities inside a quantum dot.

When all electronic states are extended (metallic regime), the random matrix
theory (RMT) \cite{Mehta} predicts the Porter-Thomas distribution of
eigenfunction intensity that only depends on the fundamental symmetry of the
quantum dot: ${\cal P}\left( v\right) =\frac{\beta /2}{\Gamma \left( \frac 
\beta 2\right) }\left( \frac{\beta v}2\right) ^{\beta /2-1}\exp \left(
-\beta v/2\right) $. The parameter $\beta =1$ for a system having
time-reversal symmetry (orthogonal ensemble), whereas $\beta =2$ when
time-reversal symmetry is broken (unitary ensemble), and $\beta =4$ for a
system having time-reversal symmetry and strong spin-orbit interactions
(symplectic ensemble). The RMT predictions are known to be valid in the
limit of large conductance $g=E_c/\Delta \gg 1$ (here, $E_c=\hbar D/L^2$ is
the Thouless energy, $D$ is the classical diffusion constant, $L$ is the
system size, and $\Delta $ is the mean level spacing) providing the system
size $L$ is much larger than the electron mean free path $l$. Corrections to
the Porter-Thomas distribution calculated in the framework of the $\sigma $%
-model formalism \cite{FM} are of order $g^{-1}$ in the weak-localization
domain.

The limiting case of orthogonal symmetry corresponds to systems with pure
potential electron-impurity scattering, while the case of unitary symmetry
applies to systems in a strong{\em \ }magnetic field that breaks the
time-reversal symmetry completely. For intermediate magnetic fields, a
crossover occurs between pure orthogonal and pure unitary symmetry classes.
This crossover in the distribution function ${\cal P}\left( v\right) $ was
previously studied within the framework of supersymmetry techniques \cite{FE}
and within an approach \cite{Kogan} exploiting the analogy with the
statistics of radiation in the regime of the crossover between ballistic and
diffusive transport.

We consider below the problem of eigenfunction statistics of chaotic
electrons in a quantum dot in an arbitrary magnetic field in the framework
of RMT. Our treatment is related to the case of the metallic regime, and
describes statistical properties of eigenfunctions amplitudes {\em and}
phases in the regime of the orthogonal-unitary crossover. The results are
applied to describe distributions of level widths and conductance peaks for
weakly disordered quantum dots in the Coulomb blockade regime in the
presence of an arbitrary magnetic field. \newpage\ 

\section{Distribution of local amplitudes and phases of eigenfunctions}

In order to study the statistical properties of electron eigenfunctions
within the RMT approach, we replace the microscopic Hamiltonian ${\cal H}$
of an electron confined in a dot by the $N\times N$ random matrix ${\bf H}%
=S_\beta \widehat{\varepsilon }S_\beta ^{-1}$ that exactly reproduces the
energy levels $\varepsilon _n$ of the electron in a dot for a given impurity
configuration. Here, $\widehat{\varepsilon }=$diag$\left( \varepsilon
_1,...,\varepsilon _N\right) $, $S_\beta $ is a matrix that diagonalizes
matrix ${\bf H}$ (parameter $\beta $ reflects the system symmetry), and it
is implied that $N\rightarrow \infty $. An ensemble of such random matrices
reproduces the electron eigenlevels for microscopically different but
macroscopically identical realizations of the random potential, and
therefore it should describe the level statistics of electrons inside the
dot. As long as we consider the metallic regime, the relevant ensemble of
random matrices is known to belong to the Wigner-Dyson class \cite
{Mehta,Brezin-Zee}. Correspondingly, averaging over randomness of the system
is replaced by averaging over the distribution function $P\left[ {\bf H}%
\right] \propto \exp \left\{ -\text{tr}V\left[ {\bf H}\right] \right\} $ of
the random-matrix elements, where $V\left( \varepsilon \right) $ is a
so-called ``confinement potential'' that grows at least as fast as $\left|
\varepsilon \right| $ at infinity \cite{Weidenmuller,FKY}. In such a
treatment, the columns of a diagonalizing matrix $S_\beta $ contain
eigenvectors of the matrix ${\bf H}$, that is $\psi _j\left( {\bf r}%
_i\right) =\left( N/V\right) ^{1/2}\left( S_\beta \right) _{ij}$, provided
the space inside a dot is divided onto $N$ boxes with radius vectors ${\bf r}%
_i$ enumerated from $1$ to $N$ (the coefficient $\left( N/V\right) ^{1/2}$
is fixed by the normalization condition $\int_Vd{\bf r}\left| \psi _j\left( 
{\bf r}\right) \right| ^2=1$).

For the case of pure orthogonal symmetry, the eigenfunctions are real, being
the columns of the orthogonal matrix $S_1$ (up to a normalization constant $%
\left( N/V\right) ^{1/2}$). When time-reversal symmetry is completely
broken, the eigenfunctions are complex and may be thought of as elements of
the unitary matrix $S_2$. We note that in this case, the real and imaginary
parts of the eigenfunction are statistically independent, and $\left\langle
\left| \text{Re}\psi _j\left( {\bf r}\right) \right| ^2\right\rangle
=\left\langle \left| \text{Im}\psi _j\left( {\bf r}\right) \right|
^2\right\rangle $. It is natural to assume that in the transition region
between orthogonal and unitary symmetry classes, the eigenfunctions of
electrons can be constructed as a sum of two independent (real and
imaginary) parts with different weights, such as $\left\langle \left| \text{%
Im}\psi _j\left( {\bf r}\right) \right| ^2\right\rangle /\left\langle \left| 
\text{Re}\psi _j\left( {\bf r}\right) \right| ^2\right\rangle =\gamma ^2$,
where the transition parameter $\gamma $ accounts for the strength of the
symmetry breaking. This means that in the crossover region, the
eigenfunctions $\left( V/N\right) ^{1/2}\psi _j\left( {\bf r}_i\right) $ can
be imagined as columns of the matrix $S$, 
\begin{equation}
\begin{tabular}{lll}
$S=\frac 1{\sqrt{1+\gamma ^2}}\left( O+i\gamma \widetilde{O}\right) $, & $%
OO^T=1$, & $\widetilde{O}\widetilde{O}^T=1$,
\end{tabular}
\label{eq.03}
\end{equation}
composed of two independent orthogonal matrices $O$ and $\widetilde{O}$. The
parameter $\gamma $ in the parametrization Eq. (\ref{eq.03}) governs the
crossover between pure orthogonal symmetry $\left( \gamma =0\right) $ and
pure unitary symmetry $\left( \gamma =1\right) $. The values $0<\gamma <1$
correspond to the transition region between the two symmetry classes. From a
microscopic point of view, this parameter is connected to an external
magnetic field. (This issue will be discussed below.)

From Eq. (\ref{eq.03}), we obtain the moments 
\[
\mu _p\left( \gamma ,N\right) =\left\langle \left| S_{ij}\right|
^{2p}\right\rangle _S=\frac 1{\left( 1+\gamma ^2\right) ^p} 
\]
\begin{equation}
\times \sum_{q=0}^p\left( 
\begin{array}{l}
p \\ 
q
\end{array}
\right) \gamma ^{2\left( p-q\right) }\left\langle \left| O_{ij}\right|
^{2q}\right\rangle _O\left\langle \left| \widetilde{O}_{ij}\right| ^{2\left(
p-q\right) }\right\rangle _{\widetilde{O}}\text{.}  \label{eq.03a}
\end{equation}
Here $\left\langle ...\right\rangle _O$ stands for integration over the
orthogonal group \cite{Ullah}. The summation in Eq. (\ref{eq.03a}) can be
carried out using the large-$N$ formula 
\begin{equation}
\left\langle \left| O_{ij}\right| ^{2p}\right\rangle _O=\left( \frac 2N%
\right) ^p\frac{\Gamma \left( p+\frac 12\right) }{\sqrt{\pi }},
\label{eq.03b}
\end{equation}
leading to 
\begin{equation}
\mu _p\left( \gamma ,N\right) =\frac{p!}{N^p}\frac{P_p\left( X\right) }{X^p}%
\text{, }X=\frac 12\left( \gamma +\gamma ^{-1}\right) ,  \label{eq.04a}
\end{equation}
which is valid over the whole transition region $0\leq \gamma \leq 1$ in the
thermodynamic limit $N\rightarrow \infty $. (Here $P_p\left( X\right) $
denotes the Legendre polynomial). One can easily see from the properties of
Legendre polynomials that the ansatz introduced by Eq. (\ref{eq.03}) leads
to the correct moments in the both limiting cases $\gamma =0$ [$\mu _p\left(
0,N\right) =\left( 2p-1\right) !!/N^p$] and $\gamma =1$ [$\mu _p\left(
1,N\right) =p!/N^p$].

Using the RMT mapping described above, we can write the distribution
function ${\cal P}\left( v\right) $, Eq. (\ref{eq.01}), in the form 
\begin{equation}
{\cal P}\left( v\right) =\int \frac{d\omega }{2\pi }\exp \left( -i\omega
v\right) \sum_{p=0}^\infty \frac{\left( i\omega N\right) ^p}{\Gamma \left(
p+1\right) }\left\langle \left| S_{rn}\right| ^{2p}\right\rangle _S\text{,}
\label{eq.05}
\end{equation}
whence we get, with the help of Eq. (\ref{eq.04a}) 
\begin{equation}
{\cal P}\left( v\right) =X\exp \left( -vX^2\right) I_0\left( vX\sqrt{X^2-1}%
\right) \text{,}  \label{eq.10}
\end{equation}
where $I_0$ is the modified Bessel function. Equation (\ref{eq.10}) gives
the distribution function of the local amplitudes of electron eigenfunctions
in a quantum dot [see Fig. 1]. It is easy to confirm that for pure
orthogonal $\left( X\rightarrow \infty \right) $ and pure unitary $\left(
X=1\right) $ symmetries, Eq. (\ref{eq.10}) yields 
\[
\left. {\cal P}\left( v\right) \right| _{X\rightarrow \infty }=\frac 1{\sqrt{%
2\pi v}}\exp \left( -\frac v2\right) , 
\]
\begin{equation}
\;\left. {\cal P}\left( v\right) \right| _{X=1}=\exp \left( -v\right) .
\label{eq.18}
\end{equation}

A different approach to the issue of eigenfunction statistics was recently
proposed in Ref. \cite{Kogan}, whose authors obtained a single-integral
representations for ${\cal P}\left( v\right) $ [see their Eq. (17)]. It can
be shown that our Eq. (\ref{eq.10}) coincides with Eq. (17) of the cited
work provided the parameter $X$ is related to the parameter $Y$ appearing in
Ref. \cite{Kogan} by $X=\sqrt{1+1/Y}$.

The advantage of the random-matrix approach is that it allows one to
calculate in a rather simple way, the distribution functions for many other
quantities, in particular, the phase distribution given by Eq. (\ref{eq.02}).

Since within the framework of the proposed random-matrix model $\sin \varphi
_j\left( {\bf r}\right) =\gamma \widetilde{O}_{ij}/\sqrt{O_{ij}^2+\gamma ^2%
\widetilde{O}_{ij}^2}$, and $\left\langle \sin ^{2p+1}\varphi _j\left( {\bf r%
}\right) \right\rangle _{O,\widetilde{O}\ }=0$, we have to compute 
\begin{equation}
{\cal Q}\left( s\right) =\int \frac{d\omega }{2\pi }\exp \left( -i\omega
s\right) \sum_{p=0}^\infty \frac{\left( -1\right) ^p\left( \gamma \omega
\right) ^{2p}}{\Gamma \left( 2p+1\right) }\sigma _{2p}\left( \gamma \right) 
\text{,}  \label{eq.22}
\end{equation}
where the average 
\begin{equation}
\sigma _{2p}\left( \gamma \right) =\left\langle \frac{\widetilde{O}_{ij}^{2p}%
}{\left( O_{ij}^2+\gamma ^2\widetilde{O}_{ij}^2\right) ^p}\right\rangle _{O,%
\widetilde{O}}  \label{eq.22a}
\end{equation}
can be calculated using its integral representation, 
\[
\sigma _{2p}\left( \gamma \right) =\frac 1{\Gamma \left( p\right) } 
\]

\begin{equation}
\times \left\langle \widetilde{O}_{ij}^{2p}\int_0^\infty dxx^{p-1}\exp
\left\{ -x\left( O_{ij}^2+\gamma ^2\widetilde{O}_{ij}^2\right) \right\}
\right\rangle _{O,\widetilde{O}}\text{.}  \label{eq.22b}
\end{equation}
Expanding the exponent in the integrand of Eq. (\ref{eq.22b}) yields 
\[
\sigma _{2p}\left( \gamma \right) =\frac 1{\Gamma \left( p\right) }%
\int_0^\infty dxx^{p-1}\sum_{q=0}^\infty \frac{\left( -1\right) ^qx^q}{%
\Gamma \left( q+1\right) } 
\]

\begin{equation}
\times \sum_{k=0}^q\left( 
\begin{array}{l}
q \\ 
k
\end{array}
\right) \gamma ^{2\left( q-k\right) }\left\langle \left| O_{ij}\right|
^{2k}\right\rangle _O\left\langle \left| \widetilde{O}_{ij}\right| ^{2\left(
p+q-k\right) }\right\rangle _{\widetilde{O}}\text{.}  \label{eq.22c}
\end{equation}
Using Eq. (\ref{eq.03b}), we obtain after straightforward calculations

\begin{equation}
\sigma _{2p}\left( \gamma \right) =\frac 1\pi \int_0^\infty \frac{d\lambda }{%
\sqrt{\lambda }\left( 1+\lambda \right) \left( \gamma ^2+\lambda \right) ^p}%
\text{.}  \label{eq.22f}
\end{equation}
Equations (\ref{eq.22}) and (\ref{eq.22f}) yield the following formula for
the phase distribution function in the crossover regime: 
\begin{equation}
{\cal Q}\left( s\right) =\frac \gamma {\pi \sqrt{1-s^2}}\frac 1{\gamma
^2+s^2\left( 1-\gamma ^2\right) }\text{.}  \label{eq.25}
\end{equation}
As can be seen from Eq. (\ref{eq.25}), the limiting case of pure orthogonal
symmetry is characterized by the $\delta $-functional phase distribution 
\begin{equation}
\frac 12\sqrt{1-s^2}\left. {\cal Q}\left( s\right) \right| _{\gamma
\rightarrow 0}=\frac 12\delta \left( s\right) \text{,}  \label{eq.26a}
\end{equation}
whereas the case of pure unitary symmetry is described by the uniform
distribution 
\begin{equation}
\frac 12\sqrt{1-s^2}\left. {\cal Q}\left( s\right) \right| _{\gamma
\rightarrow 1}=\frac 1{2\pi }\text{.}  \label{eq.26}
\end{equation}
In the crossover region, the calculated phase distribution $q\left( s\right)
=\frac 12\sqrt{1-s^2}{\cal Q}\left( s\right) $ displays a smooth transition
from $\delta $-functional distribution in the case of pure orthogonal
symmetry ($X\rightarrow \infty $) to the uniform distribution in the case of
pure unitary symmetry ($X=1$), see Fig. 2.

In order to relate the phenomenological parameter $\gamma $ (or $X$)
entering the parametrization of Eq. (\ref{eq.03}) to the magnetic field
breaking time-reversal symmetry in the real microscopic problem, we compare
the distribution given by Eq. (\ref{eq.10}) with the exact distribution
derived within the framework of microscopic supersymmetry model (Eq. (13) in
Ref. \cite{FE}). Although these distributions have different analytical
forms, good numerical agreement is observed even for the tails of
distribution functions after appropriate rescaling of our phenomenological
parameter $X$ [see inset in Fig. 1, where the distribution function $\varphi
\left( \tau \right) =2\tau {\cal P}\left( \tau ^2\right) $ for different
values of parameter $X$ is plotted]. Analysis of the behavior of $\varphi
\left( \tau \right) $ in both cases in the region of small $\tau $ yields
the following relation between transition parameter $X$ and microscopic
parameter $X_m$ appearing in Ref. \cite{FE}:\widetext
\begin{equation}
X=2X_m^2\exp \left( X_m^2\right) \left\{ \Phi _1\left( X_m\right) \left[
\Lambda _1\left( X_m\right) -\Lambda _2\left( X_m\right) \right] +\Lambda
_2\left( X_m\right) -\frac{1-\Phi _1\left( X_m\right) }{^{X_m^2}}\Lambda
_1\left( X_m\right) \right\} \text{.}  \label{eq.20}
\end{equation}
Here $\Phi _1\left( X_m\right) =\left[ \exp \left\{ -X_m^2\right\}
/X_m\right] \int_0^{X_m}dy\exp \left\{ y^2\right\} $, and 
\begin{equation}
\Lambda _n\left( X_m\right) =\frac{\sqrt{\pi }\left( 2n-1\right) !!}{%
2^{n+1}X_m^{2n+1}}\left\{ 1-%
%TCIMACRO{\limfunc{erf} }
%BeginExpansion
\mathop{\rm erf}
%EndExpansion
\left( X_m\right) +\frac{\exp \left\{ -X_m^2\right\} }{\sqrt{\pi }}%
\sum_{k=0}^{n-1}\frac{2^{n-k}X_m^{2n-2k-1}}{(2n-2k-1)!!}\right\} \text{.}
\label{eq.21}
\end{equation}
\narrowtext\noindent
In accordance with Ref. \cite{FE}, the microscopic parameter $X_m=\left|
\phi /\phi _0\right| \left( \alpha _gE_c/\Delta \right) ^{1/2}$, where $%
\alpha _g$ is a factor depending on the sample geometry, $\phi $ is the
magnetic field flux penetrating into the cross-sectional area of a sample,
and $\phi _0$ is the flux quantum. It can be seen from Eq. (\ref{eq.20})
that for very weak breaking of time-reversal symmetry $\left( X_m\ll
1\right) $, $X\approx 2\sqrt{\pi }/3X_m$. In the opposite limit of weak
deviation from unitary symmetry $\left( X_m\gg 1\right) $, $X\approx
1+1/2X_m^2$.

The connection between parameters $X$ and $X_m$ given by Eqs. (\ref{eq.20})
and (\ref{eq.21}) allows to use the simple random-matrix model proposed here
for describing various phenomena occurring in quantum dots in arbitrary
magnetic fields.

\section{Statistics of resonance conductance of a quantum dot}

The issue of eigenfunction statistics is closely connected to the problem of
conductance of a quantum dot that is weakly coupled to external leads in the
Coulomb blockade regime. At low temperatures, the conductance of a dot
exhibits sharp peaks as a function of the external gate voltage. The height
of conductance peaks strongly fluctuate since the coupling to the leads
depends on the fluctuating magnitudes of electron eigenfunctions near the
leads. Thus far both theoretical \cite{JSA} - \cite{Alhassid} and
experimental \cite{Chang,Folk} studies were restricted to the pure symmetry
classes (with conserved or completely broken time-reversal symmetry) even in
the simplest case of two pointlike leads. Here we present the analytical
treatment of the problem for arbitrary magnetic fields.

Let us consider non-interacting electrons confined in a quantum dot of
volume $V$ with weak volume disorder (metallic regime). The system is probed
by two pointlike leads weakly coupled to the dot at the points ${\bf r}_L$
and ${\bf r}_R$. The Hamiltonian of the problem can be written in the form 
\cite{Iida} 
\[
{\cal H}=\frac{\hbar ^2}{2m}\left( i\nabla +\frac e{c\hbar }{\bf A}\right)
^2+U\left( {\bf r}\right) 
\]
\begin{equation}
+\frac i{\tau _H}V\left[ \alpha _R\delta \left( {\bf r-r}_R\right) +\alpha
_L\delta \left( {\bf r-r}_L\right) \right] ,  \label{eq.40}
\end{equation}
where $U$ consists of the confinement potential and the potential
responsible for electron scattering by impurities, $\tau _H$ is the
Heisenberg time, and $\alpha _{R(L)}$ is the dimensionless coupling
parameter of the right (left) lead. We also suppose that the coupling
between the leads and the dot is extremely weak, $\alpha _{R(L)}\ll 1$, so
that the only mechanism for electron transmission through the dot is
tunneling.

It can be shown that the heights of conductance peaks are entirely
determined by the partial level widths 
\begin{equation}
\gamma _{\nu R(L)}=\frac{2\alpha _{R(L)}}{\tau _H}V\left| \psi _\nu \left( 
{\bf r}_{R(L)}\right) \right| ^2  \label{eq.44}
\end{equation}
in two temperature regimes. When $T\ll \alpha _{R(L)}\Delta $, the heights $%
g_\nu =\frac h{2e^2}G_\nu $ are given by the Breit-Wigner formula 
\begin{equation}
g_\nu =\frac{4\gamma _{\nu R}\gamma _{\nu L}}{\left( \gamma _{\nu R}+\gamma
_{\nu L}\right) ^2}.  \label{eq.45}
\end{equation}
At higher temperatures, $\alpha _{R(L)}\Delta \ll T\ll \Delta $, the heights
are determined by the Hauser-Feshbach formula 
\begin{equation}
g_\nu =\frac \pi {2T}\frac{\gamma _{\nu R}\gamma _{\nu L}}{\gamma _{\nu
R}+\gamma _{\nu L}}\text{.}  \label{eq.46}
\end{equation}

Using the distributions $P_X\left( \gamma _{\nu R(L)}\right) $ of the
partial level widths, Eqs. (\ref{eq.44}) and Eq. (\ref{eq.10}), 
\[
P_X\left( \gamma _{\nu R(L)}\right) =\frac{X\tau _H}{2\alpha _{R(L)}}\exp
\left( -\frac{X^2\tau _H}{2\alpha _{R(L)}}\gamma _{\nu R(L)}\right) 
\]
\begin{equation}
\times I_0\left( \frac{X\sqrt{X^2-1}\tau _H}{2\alpha _{R(L)}}\gamma _{\nu
R(L)}\right) \text{,}  \label{eq.47}
\end{equation}
and assuming that the eigenfunctions of electrons near the left and right
contacts are uncorrelated, that is $\left| {\bf r}_R{\bf -r}_L\right| \gg
\lambda $, we can derive analytical expressions for the distribution of the
conductance peaks heights.

\subsection{Breit-Wigner regime}

From Eqs. (\ref{eq.45}), we conclude that the distribution of the
conductance peaks is given by

\[
R_X\left( g_\nu \right) =\int_0^\infty d\gamma _{\nu R}\int_0^\infty d\gamma
_{\nu L}P_X\left( \gamma _{\nu R}\right) P_X\left( \gamma _{\nu L}\right) 
\]
\begin{equation}
\times \delta \left( g_\nu -\frac{4\gamma _{\nu R}\gamma _{\nu L}}{\left(
\gamma _{\nu R}+\gamma _{\nu L}\right) ^2}\right) .  \label{eq.48}
\end{equation}
Straightforward calculation lead to the following integral representation 
\[
R_X\left( g_\nu \right) =\frac{\Theta \left( 1-g_\nu \right) }{\sqrt{1-g_\nu 
}}\int_0^\infty d\mu \;\mu 
\]
\[
\times \left\{ \prod_{k=\left( \pm \right) }\exp \left( -X\frac{S^{\left(
k\right) }}{f^{\left( k\right) }}\mu \right) I_0\left( \frac{S^{\left(
k\right) }}{f^{\left( k\right) }}\mu \sqrt{X^2-1}\right) \right. 
\]
\begin{equation}
\left. +\prod_{k=\left( \pm \right) }\exp \left( -XS^{\left( k\right)
}f^{\left( k\right) }\mu \right) I_0\left( S^{\left( k\right) }f^{\left(
k\right) }\mu \sqrt{X^2-1}\right) \right\} ,  \label{eq.49}
\end{equation}
where 
\[
S^{\left( \pm \right) }=1\pm \sqrt{1-g_\nu }\text{,} 
\]
\[
f_{}^{\left( \pm \right) }=a\pm \sqrt{a^2-1}\text{,} 
\]
\begin{equation}
a=\frac 12\left( \sqrt{\frac{\alpha _R}{\alpha _L}}+\sqrt{\frac{\alpha _L}{%
\alpha _R}}\right) \text{.}  \label{eq.50}
\end{equation}

The integral in Eq. (\ref{eq.49}) can be calculated, yielding the
distribution function for the conductance peaks 
\[
R_X\left( g_\nu \right) =\frac X{2\pi }\frac{\Theta \left( 1-g_\nu \right) }{%
\sqrt{1-g_\nu }} 
\]
\begin{equation}
\times \sum_{i=1}^2\frac{{\bf E}\left( k_i\right) }{{\cal M}_i\sqrt{{\cal M}%
_i^2+g_\nu \left( X^2-1\right) }}\text{,}  \label{eq.51}
\end{equation}
where the following notation has been used: 
\[
{\cal M}_i=a+\left( -1\right) ^{i+1}\sqrt{1-g_\nu }\sqrt{a^2-1}\text{,} 
\]
\begin{equation}
k_i=\sqrt{X^2-1}\frac{\sqrt{g_\nu }}{\sqrt{{\cal M}_i^2+g_\nu \left(
X^2-1\right) }}\text{,}  \label{eq.52}
\end{equation}
and ${\bf E}\left( k\right) $ stands for the elliptic integral. Equation (%
\ref{eq.51}) is valid for arbitrary magnetic field flux, and it interpolates
between the two distributions corresponding to the quantum dot with
completely broken time-reversal symmetry 
\begin{equation}
R_{X=1}\left( g_\nu \right) =\frac{\Theta \left( 1-g_\nu \right) }{2\sqrt{%
1-g_\nu }}\frac{1+\left( 2-g_\nu \right) \left( a^2-1\right) }{\left[
1+g_\nu \left( a^2-1\right) \right] ^2}  \label{eq.53}
\end{equation}
and with conserved time-reversal symmetry: 
\begin{equation}
R_{X\rightarrow \infty }\left( g_\nu \right) =\frac{\Theta \left( 1-g_\nu
\right) }{\pi \sqrt{g_\nu }\sqrt{1-g_\nu }}\frac a{1+g_\nu \left(
a^2-1\right) }\text{.}  \label{eq.54}
\end{equation}

Comparing Eqs. (\ref{eq.51}) and (\ref{eq.54}), one can see that the
influence of the magnetic field is most drastic in the region of small
heights of the conductance peaks, $g_\nu \approx 0$, [Fig. 3]. Equations (%
\ref{eq.51}) yields 
\begin{equation}
R_X\left( 0\right) =\frac X2\left( 2a^2-1\right) \text{.}  \label{eq.55}
\end{equation}
Thus, we conclude that the probability density of zero-height conductance
peaks decreases from infinity to the value $\left( 2a^2-1\right) \sqrt{\pi }%
/3X_m$ at arbitrary small magnetic fluxes penetrating into the sample.

\subsection{Hauser-Feshbach regime}

When $\alpha _{R(L)}\Delta \ll T\ll \Delta $, the distribution of the
conductance peaks is given by 
\[
F_X\left( g_\nu \right) =\int_0^\infty d\gamma _{\nu R}\int_0^\infty d\gamma
_{\nu L}P_X\left( \gamma _{\nu R}\right) P_X\left( \gamma _{\nu L}\right) 
\]
\begin{equation}
\times \delta \left( g_\nu -\frac \pi {2T}\frac{\gamma _{\nu R}\gamma _{\nu
L}}{\gamma _{\nu R}+\gamma _{\nu L}}\right) .  \label{eq.56}
\end{equation}
Carrying out the double integration and using Eq. (\ref{eq.47}), we obtain
the following formula for the function $T_X\left( \xi \right) =F_X\left(
g_\nu \right) \pi \sqrt{\alpha _L\alpha _R}/2T\tau _H$, where the
dimensionless variable $\xi =2g_\nu T\tau _H/\pi \sqrt{\alpha _L\alpha _R}$:%
\widetext
\[
T_X\left( \xi \right) =\left( \frac X2\right) ^2\xi \exp \left( -aX^2\xi
\right) \int_0^\infty d\mu \left( 1+\frac 1\mu \right) ^2\exp \left[ -\frac{%
X^2}2\xi \left( \mu f^{\left( +\right) }+\frac 1{\mu f^{\left( +\right) }}%
\right) \right] 
\]
\begin{equation}
\times I_0\left( \xi \frac{X\sqrt{X^2-1}}2\left( 1+\mu \right) f^{\left(
+\right) }\right) I_0\left( \xi \frac{X\sqrt{X^2-1}}2\frac{\left( 1+\frac 1%
\mu \right) }{f^{\left( +\right) }}\right) \text{.}  \label{eq.57}
\end{equation}
\narrowtext\noindent
This equation interpolates between two limiting distributions 
\begin{equation}
T_{X=1}\left( \xi \right) =\xi \exp \left( -a\xi \right) \left[ K_0\left(
\xi \right) +aK_1\left( \xi \right) \right]  \label{eq.58}
\end{equation}
and 
\begin{equation}
T_{X\rightarrow \infty }\left( \xi \right) =\sqrt{\frac{1+a}2}\frac 1{\sqrt{%
\xi }}\exp \left( -\frac{a+1}2\xi \right)  \label{eq.59}
\end{equation}
for unitary and orthogonal symmetries, respectively.

The distribution function Eq. (\ref{eq.57}) can be represented as a series 
\[
T_X\left( \xi \right) =\frac{X^2}2\xi \exp \left( -aX^2\xi \right) 
\]
\[
\times \sum_{n=0}^\infty \sum_{m=0}^\infty \left( \frac 1{n!m!}\right)
^2\left[ \frac{X\sqrt{X^2-1}}4\xi \right] ^{2\left( n+m\right) }\left(
f^{\left( +\right) }\right) ^{2\left( n-m\right) } 
\]
\begin{equation}
\times \sum_{s=-1-2n}^{s=1+2m}\left( 
\begin{array}{l}
2\left( n+m+1\right) \\ 
2n+s+1
\end{array}
\right) \left( f^{\left( +\right) }\right) ^sK_s\left( X^2\xi \right)
\label{eq.60}
\end{equation}
that is convenient for the calculation of the distribution of conductance
peaks in the case of weak deviations from unitary symmetry. As is the case
of very low temperatures, the distribution function of the heights of the
conductance peaks given by Eq. (\ref{eq.57}) is most affected by magnetic
field in the region $g_\nu \approx 0$: 
\begin{equation}
T_X\left( 0\right) =aX,  \label{eq.61}
\end{equation}
and therefore, the probability density of the conductance peaks of zero
height immediately drops from infinity to the value $T_X\left( 0\right)
\approx 2a\sqrt{\pi }/3X_m$ when an arbitrary small magnetic field $\left(
X_m\ll 1\right) $ is applied to a system.

\section{Conclusions}

We have introduced a one-parameter random matrix model for describing the
eigenfunction statistics of chaotic electrons in a weakly disordered quantum
dot in the crossover regime between orthogonal and unitary symmetry classes.
Our treatment applies equally to the statistics of local amplitudes and
local phases of electron eigenfunctions inside a dot in the presence of an
arbitrary magnetic field. The transition parameter $X$ entering our model is
related to the microscopic parameters of the real physical problem, and
therefore the distributions calculated within the proposed random-matrix
formalism can be used for the interpretation of experiments.

This random matrix model has also been applied to describe the distribution
function of the heights of conductance peaks for a quantum dot weakly
coupled to external pointlike leads in the regime of Coulomb blockade in the
case of crossover between orthogonal and unitary symmetry. We have shown
that the magnetic field exerts a very significant influence on the
distribution of the heights of the conductance peaks in the region of small
heights. The effect of the magnetic field consists of reducing the
probability density of zero-height conductance peaks from infinity to a
finite value for arbitrary small magnetic flux.

\begin{center}
{\bf ACKNOWLEDGMENT}
\end{center}

One of the authors (E. K.) gratefully acknowledges the financial support of
The Ministry of Science and The Arts of Israel.

\widetext

Fig. 1. Distribution function $\varphi \left( \tau \right) $ for different
parameters $X$. Inset: the same distribution function calculated from Eq. (%
\ref{eq.10}) (lines) and from Eq. (13) in Ref. \cite{FE} (figures) for
parameters $X$ and $X_m$ connected by Eq. (\ref{eq.20}). Solid line: $X=2.56$%
, triangles $X_m=0.5$; dashed line: $X=1.52$, squares: $X_m=1$; dot-dashed
line: $X=1.14$, circles: $X_m=2$.

Fig. 2. Phase distribution $q\left( s\right) $. Parameters: $X=1.5$ (curve
1), $X=5$ (curve 2), $X=10$ (curve 3).

Fig. 3. Distribution of the heights of the resonance conductance peaks in
the Breit-Wigner regime, $a=1$. Parameters: $X=1.2$ (curve 1), $X=2$ (curve
2), $X=4$ (curve 3).

\end{document}